\documentclass[letterpaper,english,aps,prl,floatfix,showpacs,twocolumn,amsfonts,amssymb,superscriptaddress,longbibliography]{revtex4-2}

\usepackage[english]{babel}

\usepackage[latin9]{inputenc}
\usepackage{verbatim}
\usepackage{amsmath}
\usepackage{amssymb}
\usepackage{graphicx}
\usepackage{esint}
\usepackage{soul}
\usepackage{wrapfig}
\usepackage{chngcntr}

\makeatletter

\usepackage{bm}
\usepackage{xcolor}
\usepackage{subfigure}
\usepackage{array}

\@ifundefined{textcolor}{}{%
 \definecolor{BLACK}{gray}{0}
 \definecolor{WHITE}{gray}{1}
 \definecolor{RED}{rgb}{1,0,0}
 \definecolor{GREEN}{rgb}{0,1,0}
 \definecolor{BLUE}{rgb}{0,0,1}
 \definecolor{CYAN}{cmyk}{1,0,0,0}
 \definecolor{MAGENTA}{cmyk}{0,1,0,0}
 \definecolor{YELLOW}{cmyk}{0,0,1,0}
}

\usepackage{bm}
\usepackage{float}

\newcommand{\be}{\begin{equation}}
\newcommand{\ee}{\end{equation}}
\newcommand{\bea}{\begin{eqnarray}}
\newcommand{\eea}{\end{eqnarray}}
\newcommand{\beq}{\begin{equation}}
\newcommand{\eeq}{\end{equation}}
\def\lb{\label}

\newcommand{\bk}{\mathbf{k}}
\newcommand{\bq}{\mathbf{q}}

\newcolumntype{x}[1]{>{\centering\arraybackslash}p{#1}}

\makeatother

\def\nn{\nonumber}
\def\lb{\label}
\def\pref#1{(\ref{#1})}

\def\a{\alpha}

\def\d{\delta}
\def\g{\gamma}

\def\th{\theta}

\def\o{\omega}

\def\G{\Gamma}

\def\O{\Omega}

\def\ra{\rightarrow}

\def\bk{{\bf k}}

\def\bq{{\bf q}}

\begin{document}

\title{Terahertz-driven parametric excitation of Raman-active phonons in LaAlO$_{3}$}

\author{M. Basini}
\altaffiliation{These authors contributed equally}
\altaffiliation{Corresponding author}
\affiliation{Department of Physics, Stockholm University, Stockholm, Sweden}
\affiliation{Physics Department, ETH Z{\"u}rich, Z{\"u}rich, Switzerland}

\author{V. Unikandanunni}
\altaffiliation{These authors contributed equally}
\affiliation{Department of Physics, Stockholm University, Stockholm, Sweden}
\affiliation{Institute of Applied Physics, Bern University, Bern, Switzerland}

\author{F. Gabriele}
\altaffiliation{These authors contributed equally}
\affiliation{CNR-SPIN, Fisciano (Salerno), Italy, c/o Universit\'a di Salerno, Fisciano (Salerno), Italy}

\author{M. Cross}
\affiliation{SLAC National Accelerator Laboratory, Menlo Park, USA}

\author{A. M. Derrico}
\affiliation{Department of Physics, Temple University, Philadelphia, USA}
\affiliation{Department of Physics, University of California, Berkeley, USA}

\author{A. X. Gray}
\affiliation{Department of Physics, Temple University, Philadelphia, USA}

\author{M. C. Hoffmann}
\affiliation{SLAC National Accelerator Laboratory, Menlo Park, USA}

\author{F. Forte}
\affiliation{CNR-SPIN, Fisciano (Salerno), Italy, c/o Universit\'a di Salerno, Fisciano (Salerno), Italy}

\author{M. Cuoco}
\affiliation{CNR-SPIN, Fisciano (Salerno), Italy, c/o Universit\'a di Salerno, Fisciano (Salerno), Italy}

\author{S. Bonetti}
\affiliation{Department of Physics, Stockholm University, Stockholm, Sweden}
\affiliation{Department of Molecular Sciences and Nanosystems, Ca' Foscari University of Venice, Venice, Italy}


\begin{abstract}

Achieving parametric excitation in an oscillating physical system involves periodically adjusting one of its parameters to modulate the oscillator's natural frequency. 
This phenomenon has been observed in numerous systems within physics and engineering, profoundly transforming modern science and technology.
Despite rapid progress, the parametric control of collective excitations, such as phonons, remains a challenge while promising to generate novel and intriguing effects in a largely unexplored field.
Here, we investigate the terahertz (THz) field-induced dynamics of Raman-active phonons in the perovskite structure of LaAlO$_3$ (LAO). Utilizing intense THz pulses, we demonstrate a novel mechanism of parametric phonon excitation marked by substantial subharmonic components.
Theoretical analysis can successfully capture the hallmarks of the observed phenomena in a physical scenario with the THz field inducing a parametric coupling between the Raman mode and pairs of acoustic phonon excitations.
\end{abstract}

\maketitle





\textcolor{black}{The perovskite family of oxides (ABO${_3}$) shows a wide range of functionalities due to the strong interplay between spin, orbital, and lattice interactions. This is in particular reflected in the different structural instabilities found in these systems. An example is the oxygen octahedra tilting and distortions, which are associated with structural phase transition, in particular, with the cooling from the undistorted, high-temperature cubic phase. The idea to control structural instabilities by a coherent drive of phonons in the material was exploited in the last decades using ultrashort laser pulses, and the emergence of new phases \cite{disa2020polarizing,basini2024terahertz, sie2019ultrafast,rini2007control, nova2019metastable,li2019terahertz, subedi2017midinfrared} or the control of existing orders \cite{nova2017effective, disa2021engineering, juraschek2020phono, juraschek2021magnetic, afanasiev2021ultrafast, mashkovich2021terahertz, davies2023phononic, luo2023large} was observed. In most of these studies, the change in the material properties was primarily linked to the externally driven displacement of Raman modes, and in particular, the ones responsible for a distortion of the equilibrium crystal structure. It is therefore natural to expect that when a Raman phonon is driven by an external excitation, a larger displacement would lead to a more significant change in the material properties.
So far, the direct excitation of Raman-active modes in solids has been achieved via low-efficient, second-order processes that rely on the scattering of light, such as Impulsive Stimulated Raman Scattering (ISRS) \cite{klein2005ers} and Sum Frequency Generation (SFG) \cite{maehrlein2017terahertz}. In principle, in noncentrosymmetric materials, a more efficient, first-order process, which couples an external electric field to the electric dipole of the phonon, is also allowed.  
However, in centrosymmetric materials, Raman phonons are not dipole active, so that, even if they play a crucial role in determining the material properties, their excitation is limited by the low efficiency of the SFG and ISRS. A possibility to indirectly access a Raman mode in centrosymmetric materials is offered by Infrared Resonant Raman Scattering (IRRS) \cite{khalsa2021ultrafast}, or by Ionic Raman Scattering (IRS) \cite{maradudin1970ionic, wallis1971ionic, humphreys1972ionic, forst2011nonlinear}. In both cases, the resonant excitation of IR-active phonons mediates the Raman scattering process through the nonlinear lattice polarizability or through the coupling of phonons by the anharmonic lattice potential (IRRS and IRS, respectively). 
In their study, Neugebauer et al. \cite{neugebauer2021comparison} suggested that IRS can strongly dominate ISRS in exciting the lower energy Raman phonon (antiferrodistortive E$_g$) in the insulating perovskite (LaAlO${_3}$). It should be pointed out that the efficiency of the IRS is strictly related to the magnitude of anharmonicity of the lattice potential and to the presence of an IR-active phonon strongly coupled to the target-Raman active one. Such a coupling relies on selection rules that are hardly achievable in a general compound, so that, even if IRS is an efficient complementary excitation path to the SFG and ISRS, it suffers from a lack of generality. 
In this context, exploring new directions for the excitation of Raman-active phonons in centrosymmetric materials would open new directions for the control of the structural phases and related macroscopic properties.}

In this work, we present evidence of a yet unexplored mechanism for coherent parametric excitation of a low-energy Raman-active phonon in centrosymmetric lanthanum aluminate, (LaAlO${_3}$), which makes use of an intense THz electric field. 
We demonstrate that an intense terahertz pulse with a central frequency between 1 and 2 THz not only coherently excites the Raman-active E$_g$ phonon at 1.08 THz via ISRS and SFG, but also generates significant subharmonic spectral components, which are distinct signatures of an underlying parametric-driving mechanism.
This mechanism is based on the absorption of photons through an optical transition involving a pair of acoustic vibrational modes. The interaction between the Raman-active and acoustic modes results in a parametric excitation of the Raman-active phonon, leading to the emergence of dynamical components at subharmonic frequencies.

\begin{figure*}[t]
\centering
\includegraphics[width=16.5cm
, clip=true]{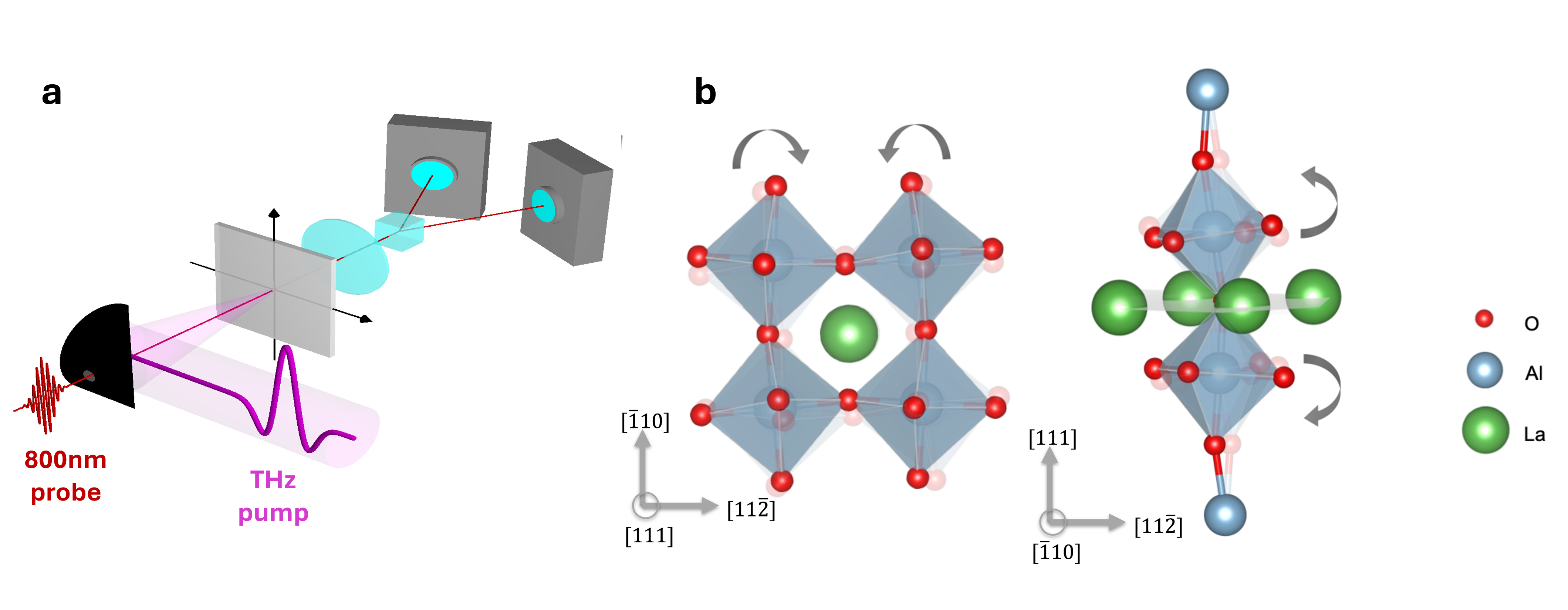}
\caption{Sketches of the experiment and of LaAlO$_{3}$ unit cell and phonon dynamics.(a) Schematic of the pump-probe setup used in this study. (b) LaAlO$_{3}$ crystal structure in the rhombohedral phase and atomic displacements involved in the E$_g$ mode} 
\label{fig1}     
\end{figure*}

\begin{figure*}[t]
\includegraphics[width=17cm,clip=true]{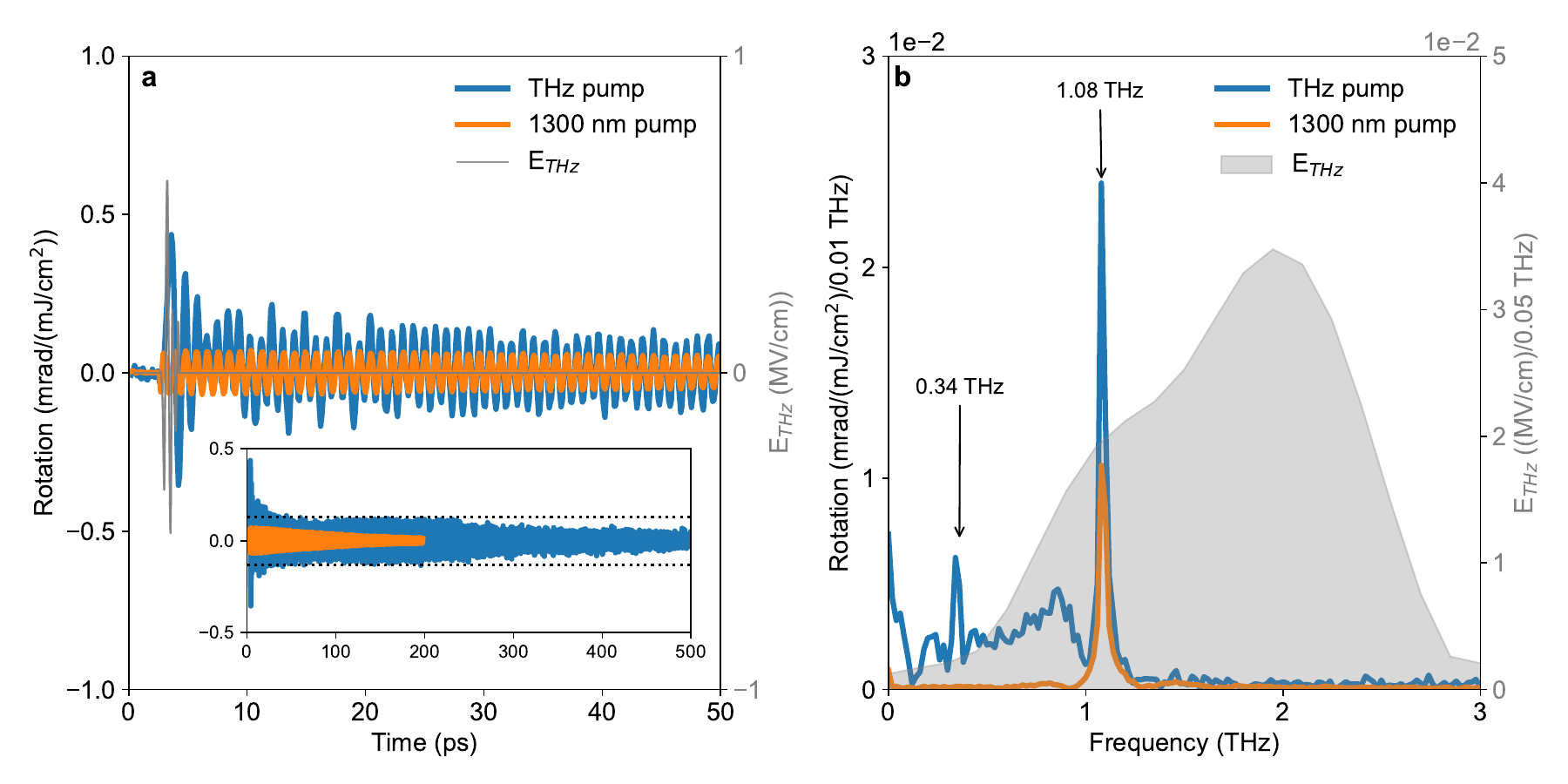}
\caption{Experimentally detected polarization rotation in LaAlO$_{3}$ following near-infrared or THz pumps.Polarization rotation of a transmitted 800 nm probe through a LaAlO$_{3}$ crystal, following the excitation by a near-infrared (1300nm) (orange curve) or a broadband THz excitation (blue curve), in (a) time and (b) frequency domains. Measurements are performed at 8 K. The FFT is evaluated in the range of 0-50 ps. The sample response is normalized by the pump fluence. Inset in panel (a): time-domain dynamics at longer timescales.}  
\label{fig2}     
\end{figure*}

LAO is a versatile material: as a single bulk crystal, it is commonly used as a substrate for the epitaxial growth of other perovskite oxides, and, as a thin film, it constitutes a building block for heterostructures and devices based on functional interfaces \cite{COLL20191}. It is a wide gap insulator ($E_{\text{gap}}$=6.1 eV), with a perovskite structure that undergoes a cubic R$\overline{3}$c to rhombohedral Pm$\overline{3}$m phase transition at $T=813$ K, associated with the rotation of the AlO$_6$ octahedra in anti-phase about the pseudocubic [111] axis \cite{hatt2010structural}. At the same temperature, the material becomes ferroelastic\cite{kustov2018laalo3,yokota2020domain}, where elastic properties and low-energy optical vibrations are intimately related\cite{carpenter2009elasticI,carpenter2009elasticII, carpenter2009elasticIII,carpenter2009elasticIV}. 
For our experiments, we use a double-sided polished LaAlO$_{3}$[100] bulk crystal with a thickness of 0.5 mm. The material's nonlinear optical response is measured in a pump-probe experimental configuration, where the pump used is either a near-infrared pulse at 1300 nm or a broadband single-cycle terahertz pulse. In Fig.\ref{fig1}(a), we show the experimental configuration. The near-infrared radiation is generated via optical parametric amplification of a 40-fs-long pulse centered at 800 nm, produced by a 1 kHz Ti:Sapphire amplifier. The pulse duration of the near-infrared pump is 60 fs. Broadband (0.5-3 THz) single-cycle terahertz radiation is generated by optical rectification of the same near-infrared laser pulse in the organic crystal OH1 \cite{brunner2008hydrogen}. The terahertz pulses are focused on the sample using parabolic mirrors to a maximum field amplitude of 700 kV/cm. We use a pair of wire-grid polarizers to control the amplitude of the THz field for conducting field-dependent measurements. The electro-optically sampled THz electric field is shown in gray in Fig.~\ref{fig2}(a) and in Fig.~\ref{fig2}(b), in time and frequency domains, respectively.
The probe pulses are produced by the same 1 kHz amplifier used to generate the pump radiation. The geometry of the pump-probe setup is shown in Fig.~\ref{fig1}(a). A half-wave plate and a Wollaston prism are used to implement a balanced detection scheme with two photodiodes.

The measurements are performed at 8 K, the temperature at which the material is centrosymmetric with a rhombohedral unit cell. Fig.~\ref{fig2} shows the signal arising from the polarization rotation of the probe beam transmitted through the sample after its excitation. Pump and probe polarizations are orthogonal to each other. To facilitate the comparison between near-infrared and THz excitations, the sample response is normalized to the incident pump fluence. A long-lasting \textcolor{black}{oscillation} is visible in Fig.~\ref{fig2}(a), which illustrates the pump-probe, time domain data for both pump wavelengths. Fig.~\ref{fig2}(b) plots the Fast Fourier Transform (FFT) of the same time-domain signal. The FFT data are dominated by a narrow peak, whose central frequency of 1.08 THz, is comparable to one of the lower-frequency Raman-active modes of LaAlO$_{3}$, i.e. the antiferrodistortive phonon mode with E$_g$ symmetry \cite{Scott1969Raman}. As illustrated in Fig.~\ref{fig1}(b), the atomic displacements of the mode are associated with the \textcolor{black}{out-of-phase} oxygen octahedra rotation about the pseudocubic axes [111].

The data presented in Fig. \ref{fig2} demonstrate that both the 1300 nm and the broadband terahertz pumps are capable of exciting the E$_{g}$ mode in the material. When excited with a broadband terahertz field, \textcolor{black}{the lifetime of the material's response is one order of magnitude longer, and it assumes a non-monotonic decay}. Moreover, additional low-frequency spectral components are visible: a quasi-continuum below 1 THz, with a broad peak around 0.8 THz and a well-defined peak at 0.3 THz. These lower frequency contributions are not observed when driving the system with the 1300 nm pump. They are too high in frequency to be ascribed to Brillouin scattering (i.e. direct excitation of acoustic modes) and are too low in frequency to be van Hove singularities of the acoustic modes, which would occur at 3 THz \cite{garrett1997vacuum}. It is worth noting that in a recent study, the broad peak at 0.8 THz was assigned to spurious effects resulting from the propagation of THz and optical probe pulses in birefringent twin domains \cite{shen2025,Sellati2025}. On the other hand, another recent investigation, monitoring the THz-driven structural dynamics of a LaAlO$_{3}$ thin film by time-resolved X-ray diffraction\cite {gollwitzer2024picosecond}, revealed the presence of a 0.3 THz mode, ruling out any implication of propagation-related effect for this component. Its origin was ascribed to a THz-driven acoustic breathing mode of the film. 
To gain a deeper insight on these features and into the symmetry of the underlying excitation mechanism, we measured the THz response of the system as a function of the probe polarizations (Fig. S4). The evidence, shown in the Supplemental Material, indicates a similar response function symmetry (i.e. Raman tensor) for both the E$_g$ mode and the low-frequency contributions. 
%
\textcolor{black}{At high temperature, i.e.120K (Figure S5), only the optical pump is still able to excite the Eg mode at 1.08 THz; the response of the THz pump showing a slow background due to the Terahertz Kerr effect. This indicates that ISRS and SFG can not be the only mechanisms responsible for the
THz-driven Eg excitation observed at 8K, and an additional mechanism, more efficient at low T, should be assumed.  Finally, the fluence dependence of the THz and optically-driven E$_g$ amplitude, presented in detail in Fig. S6 in the Supplemental Material, exhibits a linear trend, whereas the 0.34-THz mode shows a departure from linearity, displaying an additional quadratic behavior. Please note that we ruled out any effect due to local heating induced by the pump intensity, as shown in section II of the Supplementary information.}

Building on the above-presented experimental evidence, we now identify the potential excitation pathway responsible for the emergence of both the E$_g$ and the low-frequency peak in the THz-driven sample's response. 
First, we note that even if the E$_g$ mode is excited by Raman processes in the THz range (i.e., ISRS and SFG), these mechanisms alone are not sufficient to generate the observed low-frequency response. Particularly, the experimental observation of a well-defined sharp peak at $0.3$ THz, together with the absence of optical phonons and electronic excitations below 1 THz \cite{Scott1969Raman, willett2014vibrational}, points toward an additional physical scenario involving THz-field driven parametric excitation of the Raman-active phonon mode. Such a scenario can be captured by a phenomenological model, wherein light triggers two acoustic phonon modes ($Q_{ac}$) that are subsequently converted into Raman-active phonon modes ($Q_R$). \textcolor{black}{It is worth noticing that our assumption is justified by the improper ferroelestic properties of  LaAlO$_3$, implying the acoustic branch and the low-energy optical branch (the E$_g$ phonon) mixing at finite momentum, as shown in Fig.S7. This mixing leads to a strong coupling between the acoustic and E$_g$ phonon modes, first observed in \cite{hortensius2020ultrafast} and extensively discussed in \cite{li2025understanding}. Moreover, the large energy separation between the optical E$_g$ mode and the next higher-energy optical phonon (A$_g$ at 4.7 THz \cite{liu1995impulsive}) makes Raman down-conversion into acoustic phonons the dominant scattering channel for the E$_g$ mode.}

Focusing, for simplicity, on a single acoustic branch, the relevant symmetry-allowed terms in the expansion of the potential are
\begin{widetext}
\bea
V(Q_R,Q_{ac})&=&
\frac{\O_R^2}{2}
Q_R^2+
\sum_\bq\frac{\O_{ac}(\bq)^2}{2}
Q_{ac}(-\bq)Q_{ac}(\bq)
-R Q_R E_{ext}^2(t)\nonumber\\
&+& \frac{1}{2} 
\sum_\bq Z_{ac}(\bq) Q_{ac}(-\bq)Q_{ac}(\bq) E_{ext}^2(t)+\frac{1}{4}
\sum_\bq d(\bq) Q_{ac}(-\bq)Q_{ac}(\bq) Q_R^2,
\label{pot}
\eea
\end{widetext}
The first row of Eq.\ \pref{pot} describes the Raman-active and acoustic phonons oscillating at their respective frequencies, $\Omega_R$ and $\Omega_{ac}$, and the ISRS-SFG excitation mechanism for the Raman-active phonon, mediated by the Raman tensor $R$ \cite{liu1995impulsive}. The second row accounts for the simultaneous excitation of two acoustic phonons with opposite momenta by light-- allowed by both symmetry and conservation of momentum-- mediated by the effective charge $Z_{ac}$, and the anharmonic term, controlled by the coupling constant $d$, responsible for acousto-optic conversion. We emphasize that, while the Raman-active phonon couples directly to light close to the $\Gamma$ point, the full momentum dependence of the acoustic phonons and their couplings to both the Raman-active mode and light are taken into account. \textcolor{black}{As a final remark on Eq.\ \pref{pot}, terms involving the symmetry-allowed combinations $Q_{ac}^2 Q_R$ and $Q_{ac} E_{ext}^2$ have been neglected, as they are not able to account for the generation of subharmonic components.}

According to Eq.\ \pref{pot}, light directly excites the Raman-active phonon via ISRS and SFG, while simultaneously generating two acoustic phonons from the same branch, which in turn drive the system parametrically through the acousto-optic conversion mechanism sketched in Fig.~\ref{fig3}. Indeed, by integrating out the acoustic modes (see Supplemental Material), one finds that the time evolution of the Raman-active phonon is given by $Q_R(t)=R \int d t' x(t-t') E_{ext}^2(t')$, where $x(t)$ evolves as
\be
\ddot{x}(t)+
\O_R^2\left[1+f(t)\right]x(t)=0,
\label{mathieu}
\ee
with initial conditions $x(0)=0$ and $\dot{x}(0)=1$. Eq. \pref{mathieu} describes a harmonic oscillator with a time-dependent modulation of the frequency $f(t)=\left(1/\O_R\right)^2
\int d\o/\sqrt{2\pi}e^{-i\o t}
K(\o)E_{ext}^2(\o)$, which results from the convolution of the nonlinear kernel $K(\o)=\sum_\bq C(\bq)/\left[4\O_{ac}^2(\bq)-(\o+i 0^+)^2\right]$-- describing two acoustic phonons with momenta $\bq$ and $-\bq$, where $C(\bq)$ selects the dominant phonon contributions--with the squared pump field. In the following, we distinguish between narrow-band and broadband pump fields.

\begin{figure}
\includegraphics[width=0.45\textwidth]{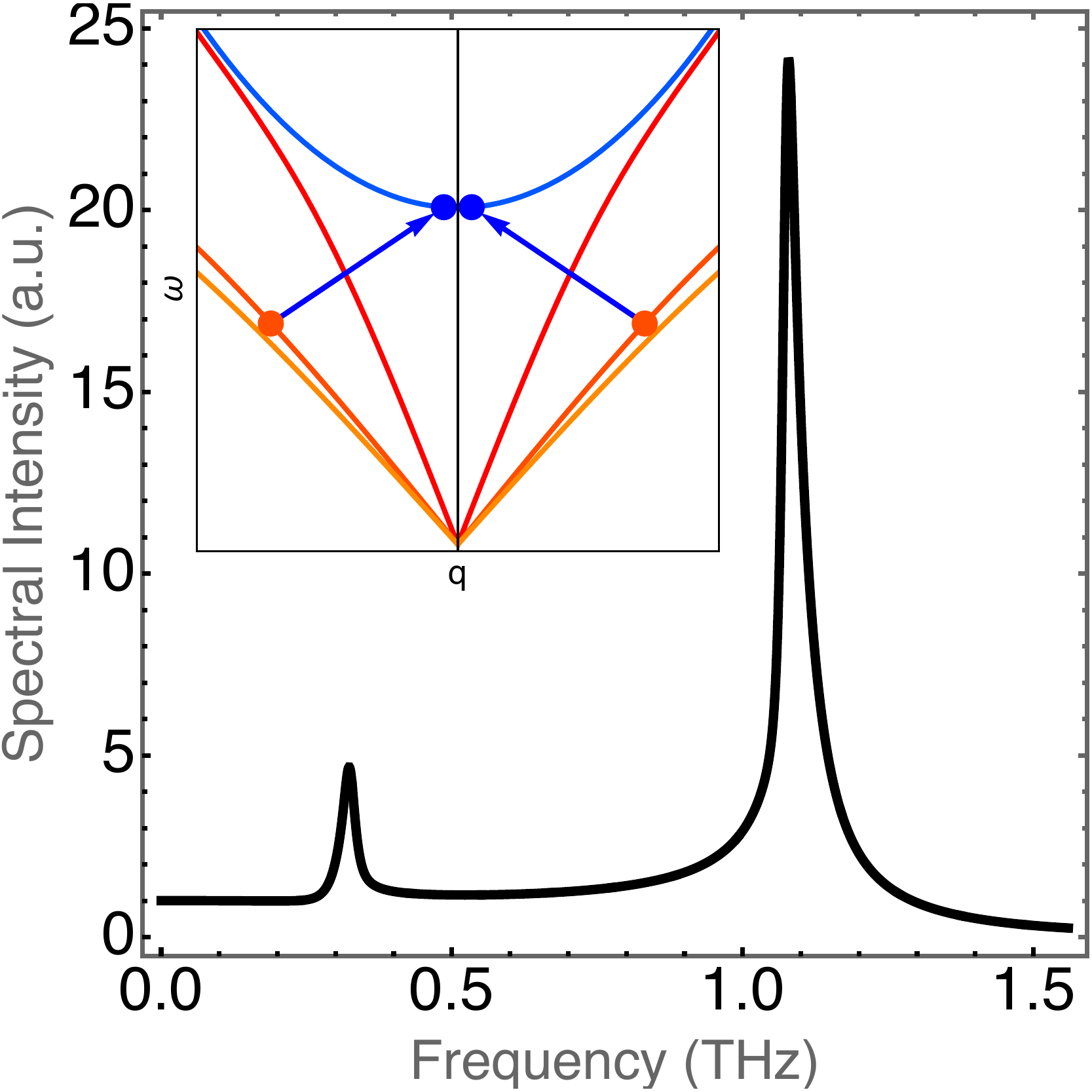}
 \centering
  \caption{Theoretical modeling of parametric phonon dynamics in LaAlO$_{3}$. Spectral intensity of the Raman amplitude $|Q_R|$ (solid black line) as a function of frequency. The inset shows the phonon dispersions: the Raman-active optical branch (blue) and three acoustic branches (shades of red). The up-conversion mechanism is illustrated, where two acoustic phonons (red dots) convert into a Raman-active phonon (blue dot).}
\label{fig3}
\end{figure}

For a narrow-band monochromatic pump field with frequency $\O_{pump}$, $f(t)=\a\cos\left(\tilde{\o} t\right)$, where $\tilde{\o}\equiv2\O_{pump}$, and Eq.\ \pref{mathieu} reduces to a Mathieu equation. When $K(2\O_{pump})\neq 0$, $\a$ is finite, and the Raman-active mode undergoes a nontrivial periodic motion that can be analyzed in the frequency domain through the Fourier transform $x(\o)=\int d t e^{i\o t} x(t)$. $x(\o)$ exhibits a prominent spectral component at the Raman characteristic frequency, accompanied by smaller components above or below $\O_R$. The emergence of subharmonic components below the characteristic frequency is a unique fingerprint of the parametric driving mechanism\cite{acar2016floquet}. To gain deeper insight into these features, we recall that a Mathieu Equation with a modulating frequency $\tilde{\o}$ defines a Floquet dynamical system with a characteristic period $T\equiv 2\pi/\tilde{\o}$ \cite{acar2016floquet}: this allows for stable solutions that can be expanded in Fourier series as
\begin{equation}
x(t)=\sum_nc_n e^{i (\O_R+n\tilde{\o}) t},
\label{floq}
\end{equation}
where a nonzero coefficient $c_n$ indicates a spectral peak at $\o=\O_R+n\tilde{\o}$. \textcolor{black}{The solution of the Mathieu equation obtained by substituting Eq.\ \pref{floq} is discussed in detail in the Supplemental Material. Here, we note that} in the non-resonant case, when  $\O_{pump}\neq\O_R$, two main regimes can be distinguished\textcolor{black}{, depending on whether $\tilde{\omega}$ lies above or below $2\Omega_R$.}. When $\tilde{\o}>2\O_R$, spectral components appear only at $\o=\O_R$ or higher frequencies. In contrast, when $\tilde{\o}<2\O_R$, nonzero spectral components appear below $\o=\O_R$; for instance, the Fourier component corresponding to $n=-1$ occurs at $\o=|\tilde{\o}-\O_R|$, \textcolor{black}{which is }a subharmonic of the main peak since $|\tilde{\o}-\O_R|<\O_R$. This analysis is general and applies whenever Eq.\ \pref{mathieu} exhibits a Mathieu-like form.

In the case of a broadband pump, such as that used in our experiment, the convolution is dominated by the characteristic frequency of the microscopic process, namely, $\tilde{\o}(\bq)=2\O_{ac}(\bq)$ for a specific momentum $\bq$. Hence, $f(t) =\sqrt{2/\pi}/\O_R^2\sum_{\mathbf{q}} C(\bq) E_{ext}^2\left(\tilde{\o}(\mathbf{q})\right)
\sin\left(\tilde{\o}(\mathbf{q}) t\right)$, with $C(\bq)$ and $E_{ext}^2\left(\tilde{\o}(\mathbf{q})\right)$ filtering the relevant momenta. 
In particular, a THz field activates smaller momenta for which $\Omega_{\text{ac}}(\mathbf{q}) < \Omega_R$, and hence $\tilde{\o}(\bq)<2\O_R$, leading to a nontrivial subharmonic spectrum. A precise comparison between theory and experiment, particularly for the broadband THz field illustrated in Fig.~\ref{fig2}, requires the full computation of $f(t)$, which in turn necessitates accounting for the momentum dependence of $Z_{\text{ac}}$ and $d$, whose computation is beyond the scope of the present work. Nevertheless, the experimental observation of a subharmonic peak at $\sim 0.3$ THz, supports the occurrence of a Mathieu-like modulation at a single frequency. Such a modulation is microscopically offered by pairs of acoustic phonons with a specific frequency, that parametrically drive the optical phonon. It is worth noting that the proposed scenario is supported by the calculated LaAlO$_3$ phonon dispersion shown in \textcolor{black}{Fig. S7}. Remarkably, the acousto-optic coupling is particularly enhanced between the Raman phonon and the upper acoustic branch along the $\Gamma$-X and the $\Gamma$-Z direction, within the energy range $1.5-4.0$ THz (i.e., 50-133 cm$^{-1}$, and corresponding momenta), where the two branches hybridizes. 
A pair of acoustic phonons from the upper branch, with energy and momentum within this range, can subsequently scatter into two pairs of lower-branch acoustic phonons, with individual frequencies $\O_1$ and $\O_2$ that are dictated by energy and momentum conservation. \textcolor{black}{Through this scattering channel,} each of these pairs then parametrically drives the Raman-active phonon. Assuming an isotropic acoustic-phonon dispersion in momentum space, with sound velocities $c_L\sim 10$km$\cdot$s$^{-1}$ and $c_T\sim 6$km$\cdot$s$^{-1}$ for the longitudinal and two degenerate transverse branches\cite{elias2018}, respectively, we consider all possible scattering processes, as detailed in the Supplemental Material. From this analysis, we find that the resulting $\O_1$ and $\O_2$ values are distributed around an average frequency $\overline{\O}\sim 0.7$ THz. Accordingly, we expand $f(t)$ around $\tilde{\omega}(\bq)= 2\overline{\O}$, so that Eq.\ \pref{mathieu} reduces, at leading order, to
\be
\ddot{x}(t)+
\O_R^2\left[1+\a\sin(
2\overline{\O} t)\right]x(t)=0,
\lb{twopar}
\ee
where $\a=\sqrt{2/\pi}C_0 E_{ext}^2(2\overline{\O})/\O_R^2$, with $C_0$ a constant. \textcolor{black}{Eq.\ \pref{twopar} is solved by substituting Eq.\ \pref{floq}, with $\a=0.2$ and an additional damping term $2\g\dot{x}(t)$, where $\g=0.01$ is evaluated from the decay of oscillations at $\o=\O_R$ in the time domain.} 
In order to compare the theoretical predictions with the experiment, we show the spectral amplitude $|Q_R(\o)|\equiv | x(\o) E_{ext}^2(\o)|$ of the Raman-active phonon as a function of $\o$ in Fig.~\ref{fig3}. \textcolor{black}{Beyond the main peak at $\O_R$, associated with a finite Fourier component $c_0$,} Eq.\ \pref{mathieu} predicts, without any fine-tuning, the emergence of a subharmonic peak\textcolor{black}{, associated with a finite $c_{-1}$, oscillating} at $\omega_{-1}\equiv2\overline{\O}-\O_R\sim 0.32$ THz, in very good agreement with the experimental findings. \textcolor{black}{To further validate the consistency of our model with the observations, we report theoretical estimates for the pump-field dependence of the subharmonic component, which are discussed in detail in the Supplementary Material. While $c_0$ is independent of the pump field, implying a quadratic pump-field dependence of the main peak (i.e., linear in the pump fluence), the subharmonic coefficient scales as $c_{-1}\sim\a$, with $\a$ proportional to the squared pump field. Therefore, the Raman amplitude at the subharmonic frequency scales as $|Q_R(\omega_{-1})|\sim|E_{ext}|^4$. This quartic behavior (i.e., quadratic in the fluence) is confirmed by the experimental pump-fluence dependence of the $0.34$-THz mode amplitude shown in Fig. S6(a) of the Supplemental Material.}
As a final remark, light could, in principle, excite multiple pairs of acoustic phonons, introducing additional parametric excitation channels and leading to a richer subharmonic spectrum. For simplicity, we have focused on the leading-order two-phonon process, which accurately captures the position \textcolor{black}{and the pump-fluence dependence} of the dominant subharmonic peak.

In conclusion, we have presented evidence for the occurrence of phononic parametric driving of a Raman-active mode through the coupling of intense terahertz pulses with acoustic waves. The resulting form of acoustic-optical interaction can explain a physical scenario in which the Raman mode is coherently excited and accompanied by the creation of dynamical components at subharmonic frequencies. Specifically, the ability to induce oscillating modes at frequencies lower than that of the Raman mode is a defining characteristic of the observed phenomenon. \textcolor{black}{It is worth noticing that in the present work, we focused on the description of the origin of the subharmonic feature. The physical origin of long lifetime and non-monotonous decay of the THz-driven response will be the topic of a future study, focused on strain-assisted long living states, as recently suggested by \cite{li2025understanding}.}
\textcolor{black}{The implications of our findings are generally vast and carry broad significance for the ultrafast, condensed-matter, and materials-science communities, stimulating further exploration of THz-driven control strategies for phonon-based functionalities. We would like to emphasise that LaAlO$_3$, being a widely used substrate and an essential building block for oxide heterostructures, offers a natural platform to exploit the mechanism introduced here in a nonlocal fashion. In particular, the parametrically excited phonon mode could be coupled to other quasiparticle excitations in an overlying film or at an interface, enabling THz control pathways that are not accessible through conventional driving schemes.}

\noindent{\bf{Acknowledgements}}\\
M.B., V.U., and S.B. acknowledge support from the Knut and Alice Wallenberg Foundation (Grant No. 2019.0068). MB acknowledges support from the Swiss National Science Foundation (Ambizione project, PZ00P2 216089). M.Cu. acknowledges support from the EU\textquotesingle s Horizon 2020 research and innovation program under Grant Agreement No. 964398 (SUPERGATE) and from PNRR MUR project PE0000023-NQSTI. F.F. and F.G. acknowledge support by the Italian Ministry of University and Research (MIUR), under grant PON 2020JZ5N9M. A.X.G. and A.M.D. acknowledge support from the DOE, Office of Science, Office of Basic Energy Sciences, Materials Sciences, and Engineering Division under Award No. DE-SC0024132. We thank M. Kareev and J. Chakhalian for providing the LAO samples. We gratefully acknowledge discussions with J. Johnson, M. Trigo, G. Orenstein, G. Khalsa and S.L. Johnson.

\bibliography{Literature.bib}

\newpage

\pagebreak
\onecolumngrid

\setcounter{figure}{0}
\renewcommand{\thefigure}{S\arabic{figure}}

\renewcommand{\thetable}{S\arabic{table}}

\setcounter{equation}{0}
\renewcommand{\theequation}{S\arabic{equation}}

\renewcommand{\thesubsection}{S\arabic{subsection}}
\renewcommand{\appendixname}{}

\begin{center}
\large\textbf{Supplemental Material}
\end{center}
\vspace{-0.7cm}

\section {THz Time-domain spectroscopy}

\begin{figure}[H]
\centering
\includegraphics[width=15cm, clip=true]{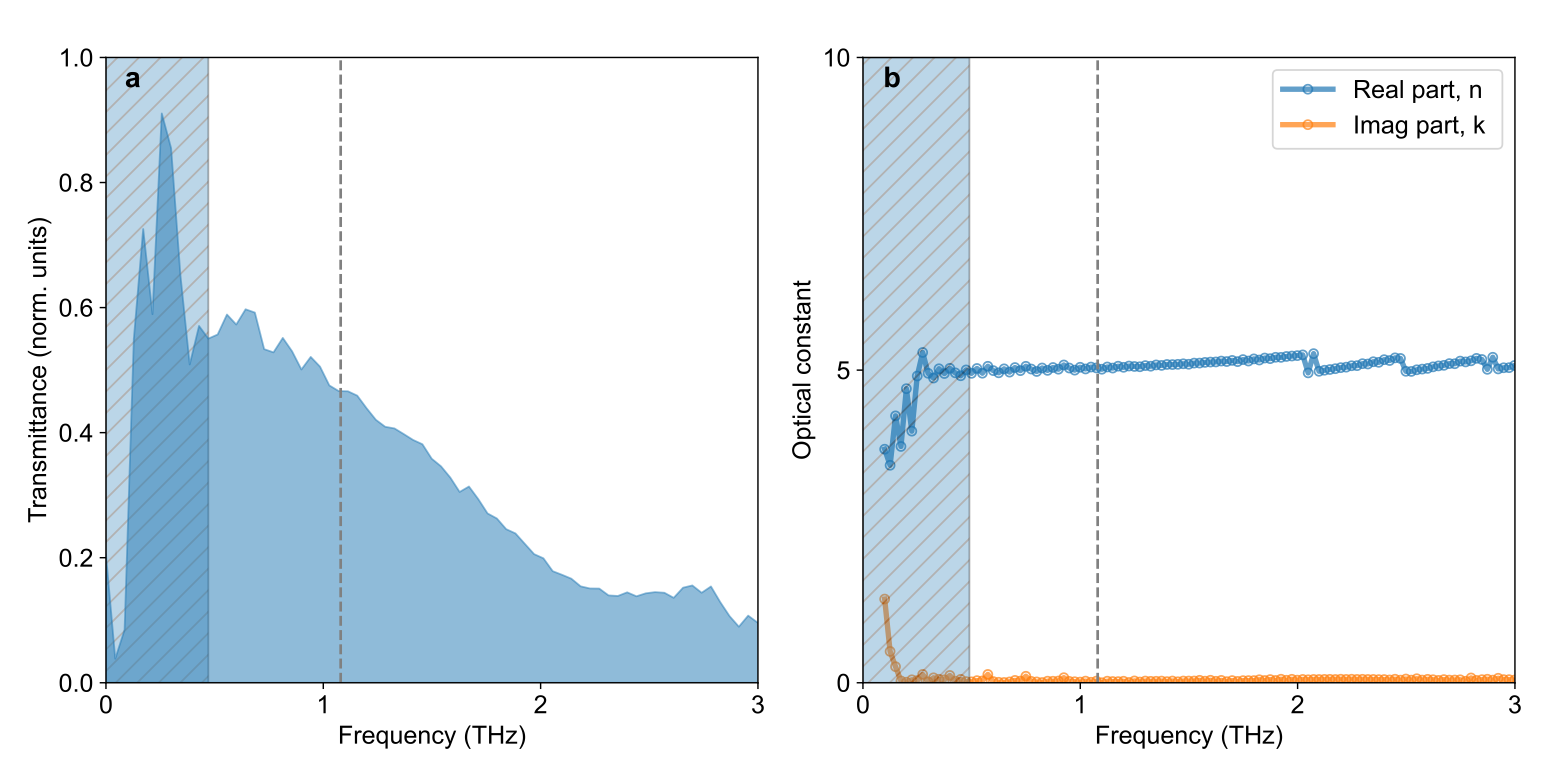}
\caption{THz-TDS on LaAlO$_{3}$ [100] sample, the incident field is the same as the pumping field of Fig.~\ref{fig2}. The field amplitude is 550kV/cm.}
\label{figS1}
\end{figure}

Here we show a static characterization (THz Time Domain Spectroscopy, THz-TDS) of the LAO sample, performed at 8K in transmission geometry. To generate a comparable field profile and amplitude to the pump-probe experiment, the incident THz field was generated in the same experimental configuration (1300nm rectification in an organic OH1 crystal). The scientific motivation is to investigate possible contributions due to the E$_g$ phonon in the complex refractive index. 
Fig.~\ref{figS1}(a) shows the sample transmittance defined as E${_T}$/E${_I}$, where E${_T}$ and E${_I}$ are the transmitted and incident fields respectively. The transmitted THz shows no absorption at 1.08 THz confirming the Raman activity of the E$_g$ mode. The calculated complex refractive index is shown in Fig.~\ref{figS1}(b) and it is consistent with previously reported data \cite{hu2004time}.
It is worth noticing that even if we show here data associated with an incident field amplitude of E${_{THz}}$=550 kV/cm, we obtained the same evidence in the amplitude range 20 kV/cm ${<}$ E${_{THz}}$ ${<}$550 kV/cm.

\textcolor{black}{\section{Local heating} In this work, we used very intense THz pulses (up to 700 kV/cm) and relatively high near-IR fluences (up to 100 mJ/cm$^{-1}$). Even if LaAlO$_{3}$ is a wide gap insulator (i.e. 6 eV) it could still suffer from cumulative heating. Here, we ruled out the role of local heating by comparing the phonon lifetime at different fluences.  If any local heating had occurred, a decrease in the phonon lifetime when increasing the fluence is expected. The plot below shows no difference in the E$_g$-driven lifetime for different pump pulse amplitudes (THz pump, Fig. S2) and energy (optical pump, Fig. S3), thus confirming that the local heating induced by both pumps is negligible.}

\begin{figure}[H]
\centering
\includegraphics[width=8cm, clip=true]{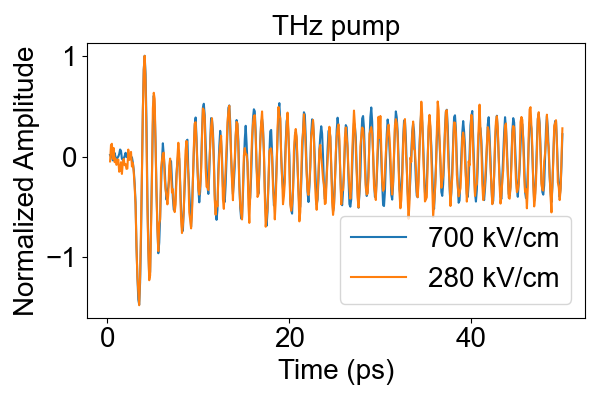}
\textcolor{black}{\caption{Comparison of the LaAlO$_3$ response to the maximum and minimum value of the electric field amplitude of the THz pump.}}
\label{figS2}
\end{figure}

\begin{figure}[H]
\centering
\includegraphics[width=8cm, clip=true]{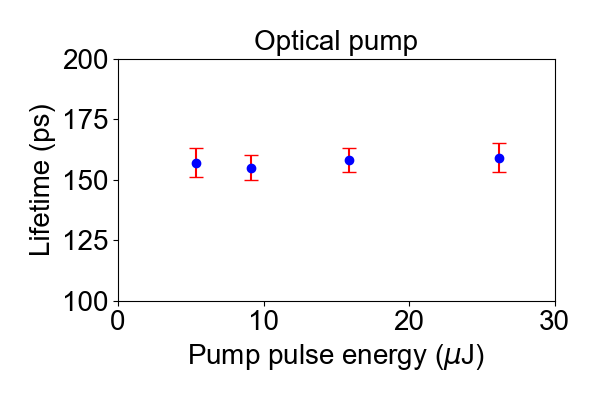}
\textcolor{black}{\caption{Comparison of time-domain fitted E$_g$ lifetime for different values of pulse energy of the optical pump. }}
\label{figS3}
\end{figure}

\section {Probe polarization dependence} 
\begin{figure}[H]
\centering
\includegraphics[width=15cm, clip=true]{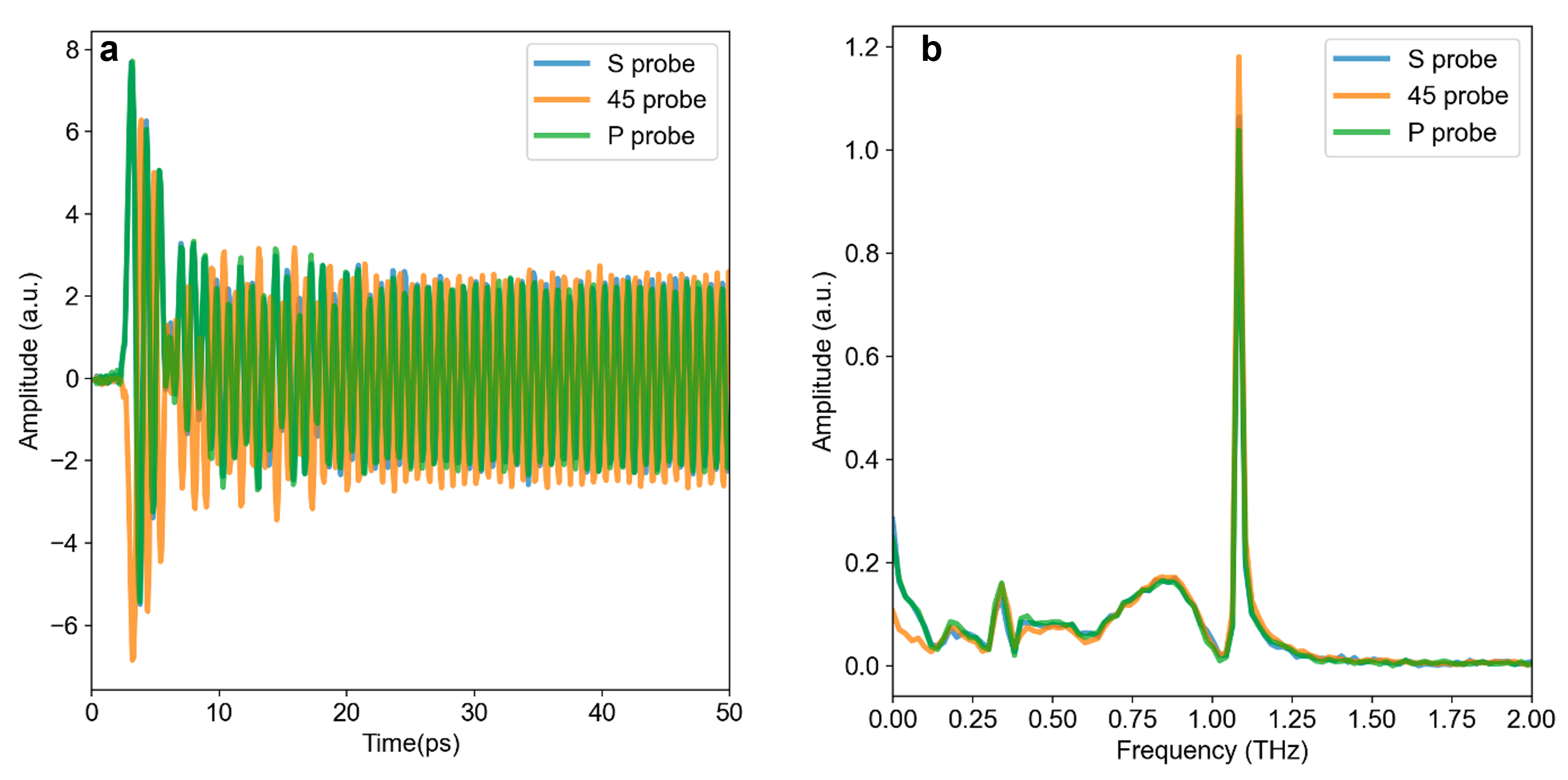}
\caption{Sample's response to broadband THz pump as a function of the incoming probe polarizations state, i.e. S (blue curve), P (green curve), and 45deg (orange curve). The rotation of the probe polarization is reported in the time domain (a) and in the frequency domain (b)}  
\label{figS4}
\end{figure}

The sample's response to different polarization states of the probe is shown in Fig.~\ref{figS4}. The relative amplitude of the 0.3 THz component with respect to the E$_g$ phonon at 1.08 THz is the same,  suggesting similar response tensors for the 0.34 THz mode and the E$_g$ mode.
\textcolor{black}{We notice that the intensity of the 0.34 THz is weaker in Fig.~\ref{figS4} than in Fig.~\ref{fig2}. We ascribe the difference to the fact that the probe polarisation dependence was performed in another spot with respect to the low-T data shown in Fig.~\ref{fig2}. LaAlO$_{3}$ is a ferroelastic material, displaying twin crystallographic domains even in its single crystal and bulk form. For this reason, we expect that the strain profile and the consequent strength of the acoustic phonon coherence may vary from spot to spot, thus leading to a change in the parametric-drive efficiency. We point out that, even if the amplitude of the 0.34 THz peak may be sensitive to the measured spot position, the experimental evidence of sub-harmonic and the overall mechanism presented in this work remains solid.}

\textcolor{black}{\section {High temperature response} 
In order to distinguish between conventional ISRS excitation, and parametric driving, we have measured the sample response at 120K to a THz and optical pump (max fluence). The evidence is shown in the figure below in the time and frequency domains and should be compared with the response at 8K shown in Figure 2 of the manuscript. At 120K, only the optical pump is still able to excite the Eg mode at 1THz by ISRS; the response of the THz pump showing a slow background due to the Terahertz Kerr effect, but no oscillation, on the other hand, at 8K both THz and optical pump were able to excite the Eg mode. This indicates that ISRS can not be the only mechanism responsible for the THz-driven Eg excitation, and an additional mechanism, more efficient at low T, should be assumed.}

\begin{figure}[H]
\centering
\includegraphics[width=15cm, clip=true]{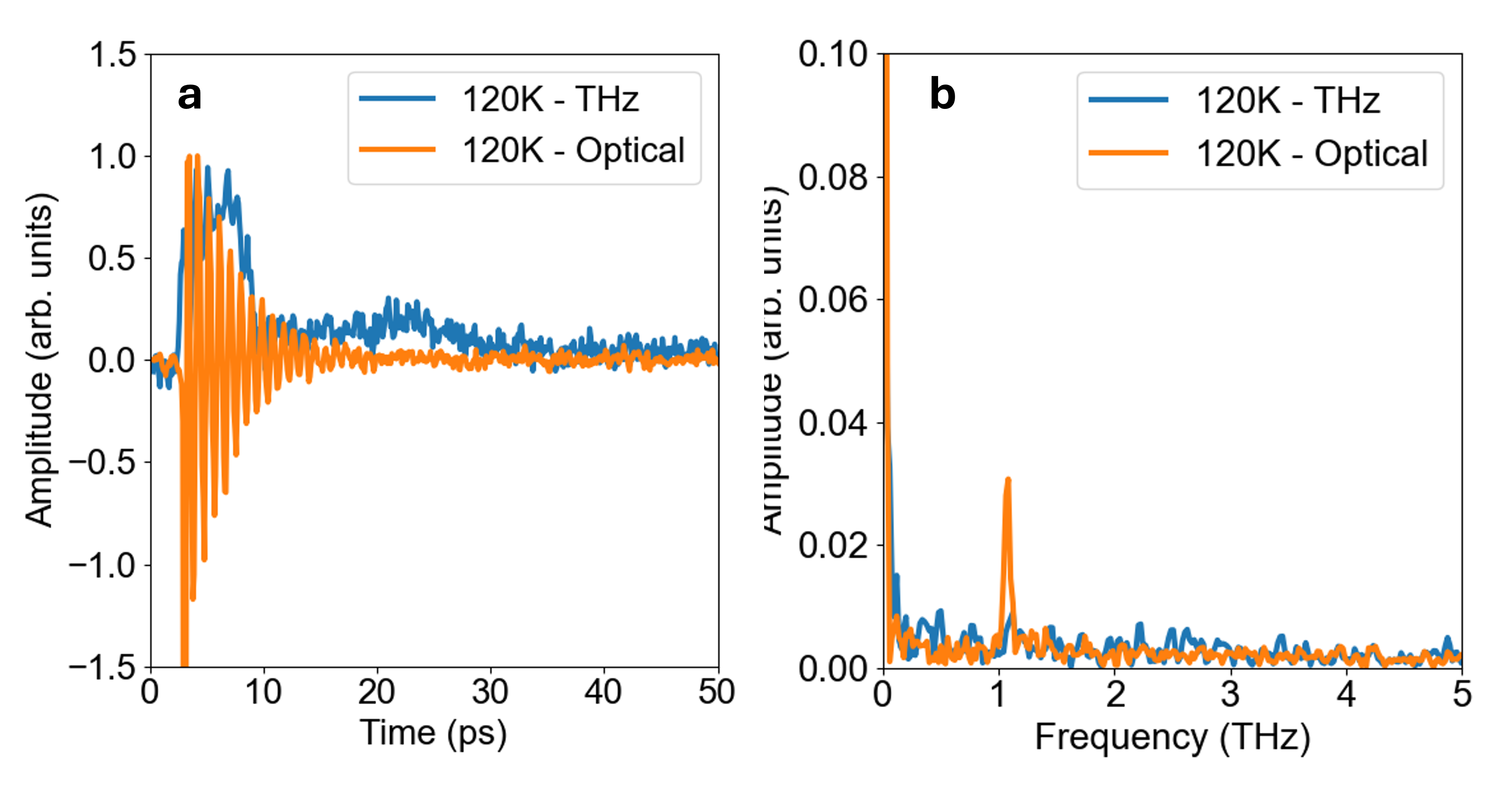}
\textcolor{black}{\caption{Normalised  LaAlO$_{3}$ response to a THz and optical pump (1300nm), in the time (a) and frequency (b) domains at 120K. }}
\label{figS3}
\end{figure}

\section{Pump fluence dependence}
 
\textcolor{black}{{Fig. S6} illustrates the pump-fluence dependence of the phonon amplitude, extracted by plotting the peak amplitudes obtained from fast Fourier transforms of the time-domain data after the THz pump pulse has fully exited the sample (time delays exceeding the THz pump round-trip time in the sample) with the pump fluence. For both pump schemes, the 
E$_g$ phonon amplitude exhibits a linear dependence on fluence, indicating a second-order process in the electric field. In contrast, the 0.34-THz component excited by the THz pulse shows a clear deviation from linearity, as shown in panel (a). Its fluence dependence reveals an additional quadratic contribution, corresponding to a fourth-order dependence on the electric field.}

\begin{figure}[!ht]
\centering

\includegraphics[width=18cm, clip=true]{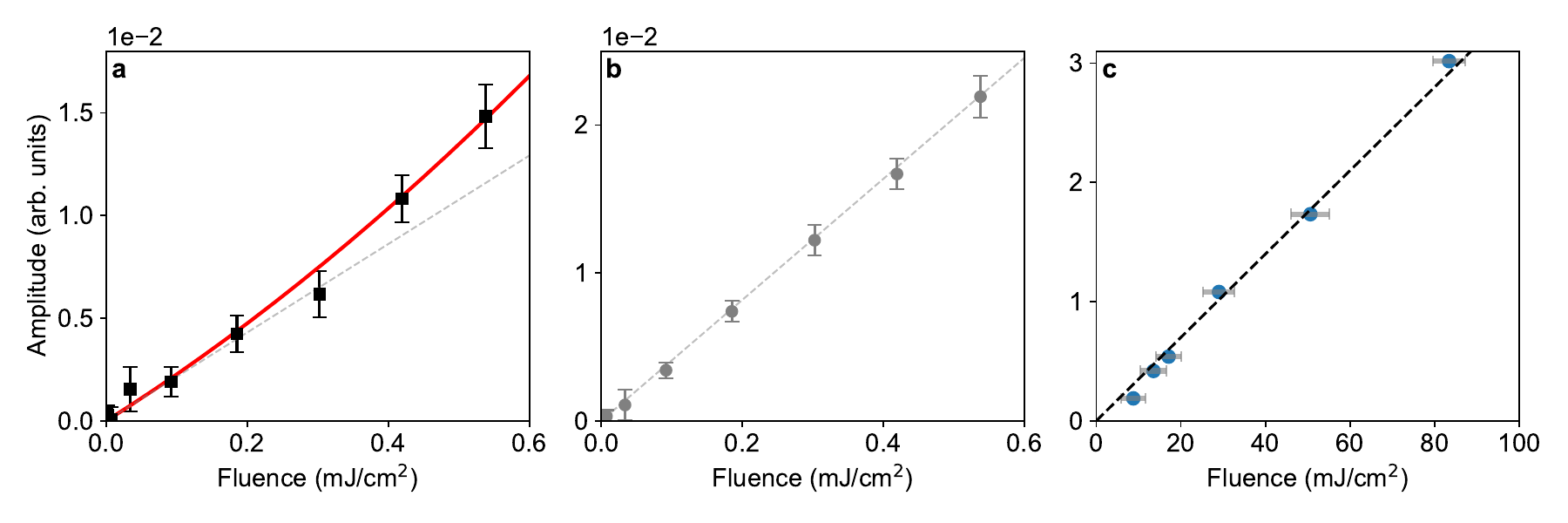}
\textcolor{black}{\caption{Pump-fluence dependence of (a) the 0.34-THz mode amplitude and (b) the E$_{g}$ phonon amplitude under THz excitation, and (c) the E$_{g}$ phonon amplitude under 1300-nm excitation. The E$_{g}$ phonon exhibits a linear fluence dependence (dashed lines) under both excitation conditions, whereas the 0.34-THz mode shows a possible departure from linearity, displaying an additional quadratic behavior as well (red continuous curve).} 
}
\label{figS5}
\end{figure}

\section {Calculated phonon dispersion}
\begin{figure}[!h]
\centering
\includegraphics[width=15cm, clip=true]{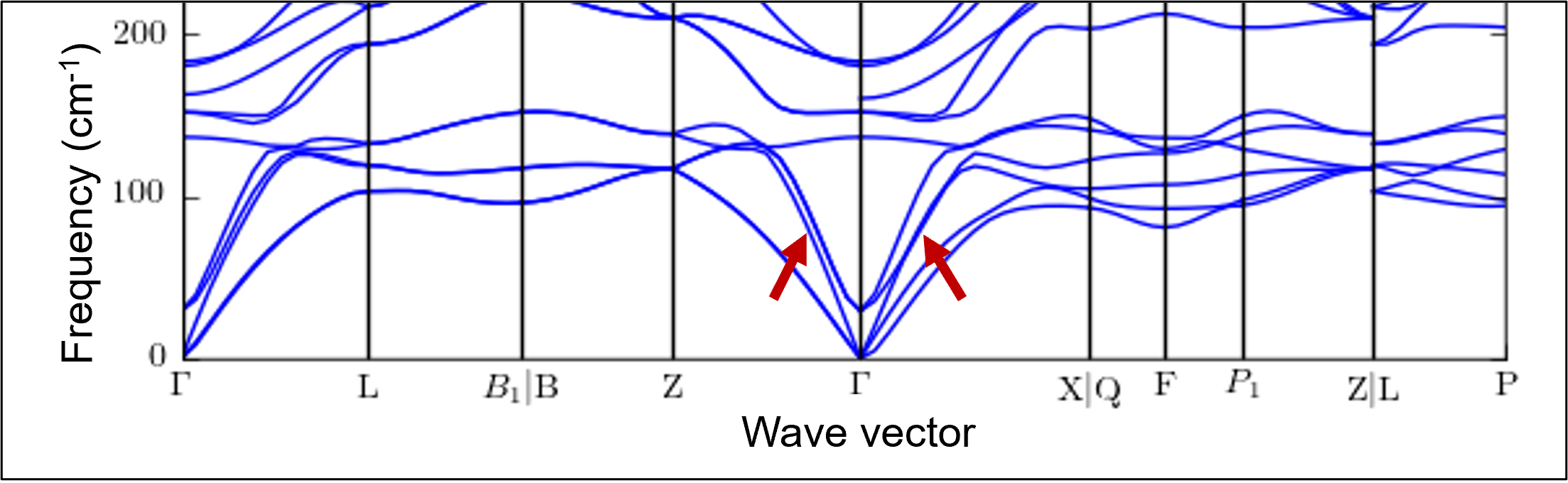}
\caption{Calculated phonon dispersion \cite{petretto2018high}.Visualization: https://legacy.materialsproject.org/materials/mp-2920/). The red arrow indicates the energy-momentum range where the acousto-optical coupling is favored.}  
\lb{figS6}
\end{figure}
The calculated phonon dispersion shown in Fig.~\ref{figS6} shows strong coupling between the lowest frequency optical phonon and the upper acoustic branches in the 50-133 cm-1 energy range, $\G$-X and $\G$-Z direction, as indicated by the red arrows.

\section{Effective Raman-active phonon dynamics}

In this section we derive the quantum correction to the quadratic action of the Raman-active phonon $Q_R$, showing that integrating out the acoustic phonons at one-loop approximation effectively results in a parametric driving of the Raman-active mode. We start from the action (with $\hbar=1$) that captures the main physical ingredients:
\be
S[Q_R,Q_{ac}]=S_G^{R}[Q_R]+S_G^{ac}[Q_{ac}]+
S_{R+ac}[Q_R,Q_{ac}]+R E_{ext}^2(t) Q_R.
\lb{total}
\ee
The first two contributions are the Gaussian actions of the Raman-active and the acoustic phonons, that read, in frequency and momentum space,
\be
S_G^R[Q_R]=\frac{1}{2}
\int d\o
\left(\o^2-\O_R^2\right)
Q_R(-\o)Q_R(\o),
\ee
\be
S_G^{ac}[Q_{ac}]=\frac{1}{2}
\sum_\bk\int d\o
\left[\o^2-\O_{ac}^2(\bk)\right]
Q_{ac}(-\bk,-\o)Q_{ac}(\bk,\o),
\ee
%
For the sake of compactness, we focus on a single acoustic branch. Also, as discussed in the main text, we neglect the momentum dependence of the Raman-active phonon frequency but account for the momentum dependence of the acoustic phonons. The term $S_{R+ac}[Q_R,Q_{ac}]$ describes the coupling of the acoustic phonons with the squared external field and the acousto-optic conversion, i.e.,
\bea
S_{R+ac}[Q_R,Q_{ac}]&=&
-\frac{1}{2}
\sum_\bk
\int \frac{d\o d\o'}{\sqrt{2\pi}}
Z_{ac}(\bk)
Q_{ac}(-\bk,-\o)
Q_{ac}(\bk,\o')
E_{ext}^2(\o-\o')\nonumber\\
&-&
\frac{1}{4}
\sum_{\bk,\bk'}
\int \frac{d\o d\o'}{\sqrt{2\pi}}
d(\bk)
Q_{ac}(-\bk,-\o)
Q_{ac}(\bk,\o') 
Q_R^2(\o-\o'),
\lb{ac+R}
\eea
where 
$Q_R^2$ stands for the convolution product $Q_R^2(\o)=
1/
\int d\o'/\sqrt{2\pi} Q_R(\o-\o')Q_R(\o')$ and $E_{ext}^2(\o)=\int d\o/\sqrt{2\pi}e^{i\o t} E_{ext}^2(t)$ is the Fourier transform of the squared pump field. The last contribution accounts for the ISRS and SFG excitation mechanism. 
%
%

\begin{figure}[t!]
\includegraphics[width=0.45\textwidth]{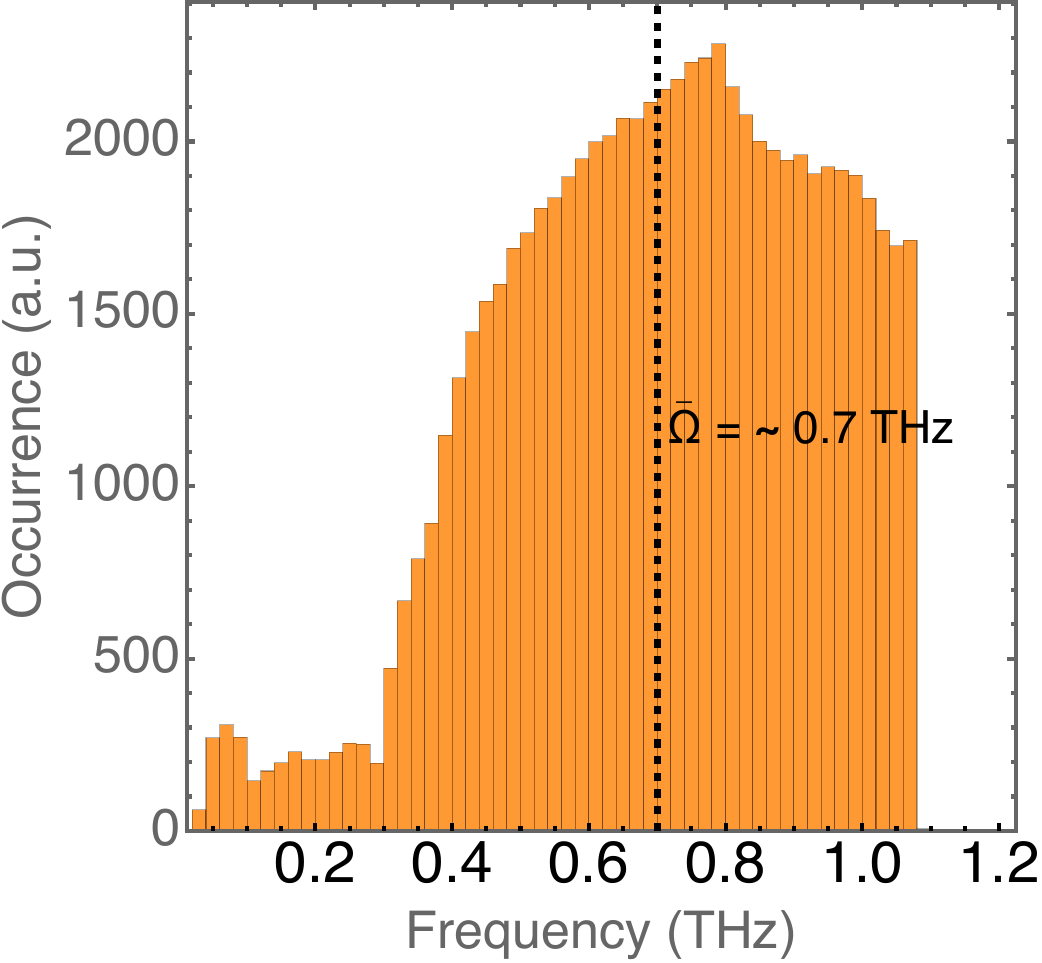}
\caption{Distribution of the frequencies of the scattered phonons in the interval $0$-$1.08$ THz relevant for the production of subharmonics via parametric driving. The dotted black line indicates the average $\overline{\O}\sim 0.7$ THz of the distribution.} 
\label{figS7} 
\end{figure}

In order to derive the effective action for $Q_R$ we notice that in the sector of the total action that contains the acoustic phonons, $S_G^{(ac)}+S_{R+ac}$, one can identify a diagonal and a non-diagonal contribution with respect to the frequency. These are, respectively,
\be
D^{-1}(\bk,\o)
\equiv\o^2-\O_{ac}^2(\bk),
\ee
i.e., the acoustic phonon propagator, and
\be
R(\bk,\o-\o')=
\frac{Z_{ac}(\bk)}{\sqrt{2\pi}} 
E_{ext}^2
(\o-\o')+
\frac{1}{\sqrt{2\pi 
}}
\frac{d(\bk)}{2}
Q_R^2(\o-\o'),
\ee
so that
\bea
&&S_G^{(ac)}[Q_{ac}]+S_{R+ac}[Q_R,Q_{ac}]\nn\\
&=&\frac{1}{2}\sum_{\bk}\int d\o \int d\o'\left[D^{-1}
(\bk,\o)\d_{\o,\o'}-R(\bk,\o-\o')\right]Q_{ac}(-\bk,-\o)Q_{ac}(\bk,\o').
\eea
By standard computations\cite{udina_prb19,gabriele2021non,fiore_preprint23,basini_prb24}, one integrates out the acoustic phonons and then makes an expansion in the non-diagonal term $R$, that yields, at one-loop approximation,
\bea
\label{corr}
\frac{\hbar}{2}
\sum_{\bk} 
\int d\o
\int \frac{d\o'}{2\pi}
D(\o+\o')D(\o')R(-\o)R(\o)
&=&\frac{1}{2}
\int d\omega Q_R^2(-\omega)
K(\o) 
E_{ext}^2(\o)\nn\\
&=&\frac{\O_R^2}{2}\int dt
f(t)
Q_R^2(t)
\eea
where we made $\hbar$ explicit again, and
\be
f(t)=
\frac{1}{\O_R^2}\int 
\frac{d\o}{\sqrt{2\pi}}
e^{-i\o t}K(\o) 
E_{ext}^2(\o),
\lb{f}
\ee
as already defined in the main text. The kernel $K$ reads, in the low-temperature limit $T\ra 0$\cite{gabriele2021non},
\be
K(\o)=
\sum_\bk
\frac{C(\bk)}{4\O_{ac}^2(\bk)-(\o+i 0^+)^2}
\label{kernel}
\ee
where $C\equiv \hbar(Z_{ac} d)/(2\O_{ac} N_\bk)$, with $N_\bk$ the effective number of sites in $\bk$-space, and $i 0^+$ is the vanishing positive imaginary part that guarantees causality. The equation of motion for the Raman-active phonon thus becomes
\be
\ddot{Q}_R(t)+\O_R^2\left[1+f(t)\right]
Q_R(t)=R E_{ext}^2(t),
\lb{effeq}
\ee
which can be solved by noting that, at long times, the solution is given by $Q_R(t)=R\int d t' x(t-t') E_{ext}^2(t')$, where $x(t)$ is the solution to Eq. \pref{mathieu} in the main text. As already discussed in the main text, for a monochromatic pump field $E_{ext}(t)=E_0\cos(\O_{pump} t)$, for which $E_{ext}^2(\o)=\sqrt{2\pi}\left(E_0/2\right)^2\left[
2\d(\o)+\d(\o-2\O_{pump})+\d(\o+2\O_{pump})
\right]$, $f(t)=\a\cos(2\O_{pump} t)$, with $\a=1/2\left(E_0/\O_R\right)^2K(2\O_{pump})$, and Eq.\ \pref{effeq} reduces to a Mathieu equation. For a generic broadband pump field, $f(t)=\sqrt{\pi/2}/\O_R^2
\sum_\bk C(\bk)/\ E_{ext}\left((2\O_{ac}(\bk)\right)\sin\left((2\O_{ac}(\bk)t\right)\th(t)$. As already discussed in the main text, $\O_{ac}$ may represent an acoustic phonon from the upper longitudinal acoustic branch, and the relevant momenta in the summation are those corresponding to the energy range in which the acousto-optical coupling is largest, i.e., $1.5-4.0$ THz. However, pairs of acoustic phonons from that branch with such individual energies cannot produce subharmonic components. Nevertheless, they can scatter into lower-energy phonons from the two transverse branches, which can then give rise to finite subharmonic components via parametric driving. Let $\O$ be the energy of a single acoustic phonon from the upper branch, with momentum $\bk$ such that $|\bk|=\O/c_L$, where $c_L$ is the sound velocity of the branch. Energy and momentum conservation ensure that it can scatter into two phonons from the lower branches, with energies $\O_1$ and $\O_2$ given by
\be
\O_1=\frac{\O}{2}\frac{1-\left(c_T/c_L\right)^2}{1-\left(c_T/c_L\right)\cos\th},
\quad
\O_2=\frac{\O}{2}
\frac{c_L^2+c_T^2-2 c_L c_T\cos\th}{c_L\left(c_L-c_T\cos\th\right)},
\ee
where $\th$ is the deflection angle of one of the two scattered phonons. Similar expressions also apply when one of the scattered phonons belongs to the upper branch. By considering all possible deflection angles and varying $\O$ uniformly over the interval $1.5-4.0$ THz, we obtain a distribution for all possible frequencies $\O_1$ and $\O_2$, as shown in Fig.~\ref{figS7}. Such a distribution has an average value of $\overline{\O}\sim 0.7$ THz and negligible higher-order moment, allowing for the expansion of $f(t)$ around $2\overline{\O}$ in Eq.\ \pref{mathieu} of the main text. 

\textcolor{black}{Lastly, we discuss the solution of the most general Mathieu equation, 
\begin{equation}
\ddot{x}(t)+\left[\O_R^2+\a\sin
\left(\tilde{\o}t\right)\right]x(t)=0,\quad\quad
x(0)=0,\quad \dot{x}(0)=1,
\lb{mathieuSup}
\end{equation}
that governs the the Raman-active phonon dynamics. Eq.\ \pref{mathieuSup} describes a Floquet dynamical system with period $T\equiv 2\pi/\tilde{\o}$. Following Ref.\ \cite{acar2016floquet}, its general solution can be written as
\begin{equation}
x(t)=e^{i\mu t}\sum_{n=-\infty}^{+\infty} c_n e^{in\tilde{\o}t},
\lb{floqSup}
\end{equation}
where $\mu$ is the Floquet exponent and the coefficients $c_n$ are nonzero for solutions of Eq.\ \pref{mathieuSup}. By direct substitution of Eq.\ \pref{floqSup} into Eq.\ \pref{mathieuSup}, we get
\begin{equation}
\left[\left(\mu+n\omega\right)^2-\O_R^2\right] c_n
-\frac{\alpha}{2i}\left( c_{n+1}-c_{n-1}\right)=0.
\lb{mathieufloq}
\end{equation}
At leading order in $\alpha$, Eq.\ \pref{mathieufloq} admits two solutions $\mu=-n_0 \tilde{\omega}\pm \O_R$, where $n_0$ is an integer. Focusing on the solution with the plus sign, substituting it into Eq.\ \pref{mathieuSup}, and shifting the index $n$ with respect to $n_0$, we find that, at leading order in $\alpha$, the solution of Eq.\ \pref{mathieuSup} can be written as
\begin{equation}
x(t)=\sum_n c_n e^{i\o_n t},
\end{equation}
with $\o_n\equiv \O_R+n\tilde{\o}$. The previous equation corresponds to Eq.\ \pref{floq} of the main text. The main peak at $\omega_0=\O_R$ is associated with $c_0=1$, while for the subharmonic peak at $\omega_{-1}=\O_R-\tilde{\omega}$ the coefficient is $c_{-1}=\frac{\alpha}{2i}\frac{1}{\left(\O_R-\tilde{\o}\right)^2-1}$.}

\end{document}